\documentclass[english,aps,prd,preprint]{revtex4}
\usepackage[T1]{fontenc}
\usepackage[latin1]{inputenc}
\usepackage{amsmath}
\usepackage{graphicx}
\usepackage{amssymb}
\usepackage{graphicx}
\usepackage{babel}

\begin{document}
\title{Single production of excited neutrinos at future $e^{+}e^{-}$, $ep$
and $pp$ colliders}
\author{O. \c{C}ak{\i}r}
\thanks{ocakir@science.ankara.edu.tr}
\affiliation{Ankara University, Faculty of Sciences, Department of Physics, 06100,
Tandogan, Ankara, Turkey }
\author{\.{I}. T\"{u}rk \c{C}ak{\i}r}
\thanks{iturk@taek.gov.tr}
\affiliation{Ankara Nuclear Research and Training Center, 06100, Tandogan, Ankara,
Turkey }
\author{Z. K{\i}rca}
\thanks{zkirca@ogu.edu.tr}
\affiliation{Osmangazi University, Faculty of Arts and Sciences, Department of
Physics, 26480, Meselik, Eskisehir, Turkey}

\begin{abstract}
We study the potential of the linear collider (LC) with
$\sqrt{s}=0.5$ TeV, linac-ring type ep collider (LC$\otimes$LHC)
with $\sqrt{s}=3.74$ TeV and the large hadron collider (LHC) with
$\sqrt{s}=14$ TeV to search for excited neutrinos through
transition magnetic type couplings with gauge bosons. The excited
neutrino signal and corresponding backgrounds are studied in
detail to obtain accessible mass limits and couplings for these
three types of colliders.
\end{abstract}


\maketitle

\section{introduction}

The proliferation of fermionic generations and their complex
pattern of masses and mixing angles are expected to be addressed
by composite models \cite{Terazawa}. A typical consequence of
compositeness is the appearance of excited leptons ($l^\star$) and
quarks ($q^\star$) \cite{Renard,Baur,Hagiwara85}. Charged
($e^{\star},\mu^{\star}$ and $\tau^{\star}$) and neutral
($\nu_{e}^{\star},\nu_{\mu}^{\star}$ and $\nu_{\tau}^{\star}$)
excited leptons are predicted by composite models where leptons
and quarks have a substructure. Phenomenologically, an excited
lepton is defined to be a heavy lepton which shares leptonic
quantum number with one of the ordinary leptons.

Current limits on the masses of excited neutrinos are
\cite{Hagiwara02}: $m_{\star}>99.4$ GeV from LEP (pair production)
assuming $f=f'$ \cite{Acciarri}, $m_{\star}>171$ GeV from LEP
(single production) assuming $f=-f'=\Lambda/m_{\star}$
\cite{Acciarri} and $m_{\star}>114$ GeV from HERA (single
production) assuming $f=-f^{'}=\Lambda/m_{\star}$ \cite{Adloff}.

The production of excited leptons was studied at LEP and HERA
energies \cite{Hagiwara85,Boudjema} and at hadron colliders
\cite{Eboli} by taking into account signal and background. The LEP
and HERA bounds on excited lepton masses are low enough when
compared to the expected scale $\Lambda$ for the compositeness.
This motivates us to reanalyze the excited lepton production at
the future colliders.

In this paper, we consider the least studied (compared to charged
excited lepton production) excited neutrino production in more
detail. We take into account signal as well as the backgrounds
(with the interference between them) at the similar experimental
conditions and compare the potential of each type of colliders to
search for the single production of excited neutrinos.

Three types of colliders related to the energy frontiers in
particle physics research seem to be promising in the next decade.
Namely, they are Large Hadron Collider (LHC) with the center of
mass energy $\sqrt{s}=14$ TeV and luminosity
$L=10^{34}$cm$^{-2}$s$^{-1}$, linear $e^{+}e^{-}$ collider (LC)
with $\sqrt{s}=0.5$ TeV and $L=10^{34}$cm$^{-2}$s$^{-1}$, and
linac-ring type ep collider (LC$\otimes$LHC) with $\sqrt{s}=3.74$
TeV and $L=10^{31}$cm$^{-2}$s$^{-1}$ (see \cite{Sultansoy} and
references therein). Even though the last one has a lower
luminosity it can provide better conditions for investigations of
a lot of phenomena comparing to LC due to the essentially higher
center of mass energy and LHC due to more clear environment. For
this reason, different phenomena (compositeness, SUSY, etc.)
should be analyzed taking into account all three types of
colliders. This work is a continuation of previous work devoted to
the study of excited electrons \cite{Cakir03,Cakir04}.

Excited leptons can be classified by SU(2)$\times$U(1) quantum
numbers as sequential type, mirror type and homodoublet type
\cite{Hagiwara02}. In the present study, we will assume that the
excited fermions have spin and isospin 1/2; higher spin and
isospin assignments have been discussed in \cite{Kuhn}. We assume
that excited lepton acquire their masses prior to SU(2)$\times$
U(1) symmetry breaking. Therefore, we consider their left- and
right-components in weak isodoublets. The success of QED
prediction for $g-2$ and small masses of leptons suggest chirality
conservation, i.e., an excited lepton should not couple to both
left- and right-handed components of the corresponding lepton.
Transition magnetic type couplings between ordinary and excited
leptons are described by the effective Lagrangian \cite{Boudjema}
\begin{equation}
L=\frac{1}{2\Lambda}\bar{l_{R}}^{\star}\sigma^{\mu\nu}
\left[fg\frac{\overrightarrow{\tau}}{2}\cdot\overrightarrow{W}_{\mu\nu}
+f^{'}g^{'}\frac{Y}{2}B_{\mu\nu}\right]l_{L}+h.c.
\end{equation}
where $\Lambda$ is the scale for the new physics that excited
leptons can appear; $W_{\mu\nu}$ and $B_{\mu\nu}$ are the field
strength tensors; $\overrightarrow{\tau}$ denotes the Pauli
matrices, $Y=-1/2$ is the hypercharge; $g$ and $g^{'}$ are the SM
gauge couplings of SU(2) and U(1), respectively; the constants $f$
and $f^{'}$ are the scaling factors for the corresponding gauge
couplings.

The effective lagrangian gives rise to the following excited
neutrino-lepton-gauge boson vertices:
\begin{eqnarray}
V_{\alpha}^{\nu^{*}\nu\gamma}
&=&\frac{g_{e}(f-f^{^{\prime}})I_{3}}
{2\Lambda}q^{\beta}\sigma_{\alpha\beta}
(1-\gamma_{5})\nonumber\\
V_{\alpha}^{\nu^{*}\nu Z} & =&\frac{g_{e}(f\cot\theta_{W}
+f^{^{\prime}}\tan\theta_{W})I_{3}}{2\Lambda}q^{\beta}\sigma_{\alpha\beta}(1-\gamma_{5})\nonumber\\
V_{\alpha}^{\nu^{*}eW} & =&\frac{g_{e}f}{2\sqrt{2}
\Lambda\sin\theta_{W}}q^{\beta}\sigma_{\alpha\beta}(1-\gamma_{5})
\end{eqnarray}

\begin{figure}
\includegraphics[scale=0.8]{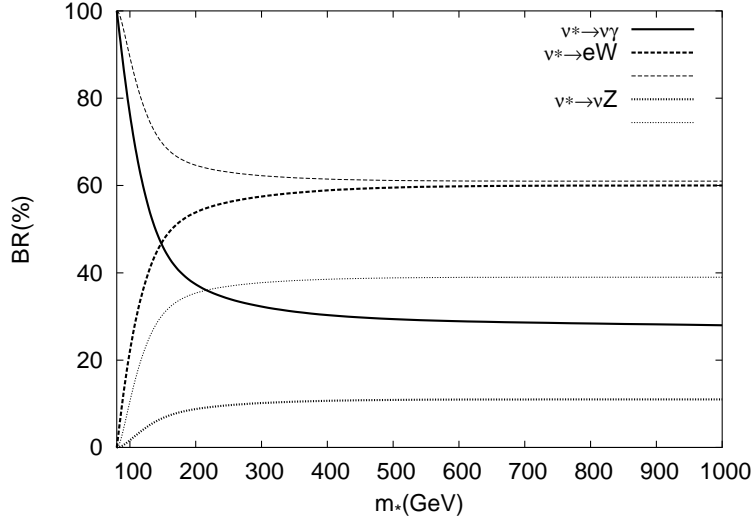}
\caption{The branching ratios BR (\%) depending on the mass of excited neutrino
for $f=f'=1$ (thick curves) and $f=-f'=1$ (thin curves). \label{fig1}}
\end{figure}

For an excited neutrino, three decay modes are possible: radiative
decays $\nu^{\star}\rightarrow \nu\gamma$, neutral currents decays
$\nu^{\star}\rightarrow \nu Z$ and charged current decays
$\nu^{\star}\rightarrow e W$. Neglecting ordinary lepton masses
the decay widths are obtained as
\begin{equation}
\Gamma(\nu^{\star}\rightarrow (e,\nu)V)=\frac{\alpha m_{\star}^{3}}{4\Lambda^{2}}
f_{V}^{2}\left(1-\frac{m_{V}^{2}}{m_{\star}^{2}}\right)^{2}\left(1+\frac{m_{V}^{2}}{2m_{\star}^{2}}\right)
\end{equation}
where $f_V\equiv (f_{\gamma}=(f-f^{'})/2$,
$f_{W}=f/(\sqrt{2}\sin\theta_{W})$ and
$f_{Z}=(f\cot\theta_{W}+f^{'}\tan\theta_{W})/2$); $m_V$ is the
mass of the gauge boson; $\alpha$ is the electromagnetic coupling
constant defined by $g_e=\sqrt{4\pi\alpha}$. The total decay width
of the excited neutrino is $\Gamma=1.0$ (6.9) GeV for
$m_\star=200$ (1000) GeV at $f=f^{'}=1$ and $\Lambda=m_\star$. The
branching ratios (BR) of excited neutrino into ordinary leptons
and gauge bosons $\gamma,Z,W$ are given in Fig. \ref{fig1}. For
large values of the excited neutrino mass, the branching ratio for
the individual decay channels reaches to the constant values: 61\%
for the $W$-channel, 39\% for the $Z$-channel while the photon
channel vanishes at $f=f^{'}=1$ . At $f=-f^{'}=1$ the radiative
decay is allowed for excited neutrino whereas it is forbidden for
excited electron. In this case, branchings will be 61\% for
$W$-channel, 11\% for $Z$-channel and 28\% for photon channel at
higher excited neutrino masses ($m_*>500$ GeV). Therefore, the
dominating signature of $\nu^\star\to W^+e^-$ is preferable for
the investigation of excited neutrino in future experiments.

\section{Single production of excited neutrinos}

We analyze the potentials of the LC, LC$\otimes$LHC and LHC
machines to search for excited neutrinos (or antineutrinos) via
the single production reactions
\begin{equation}
e^{+}e^{-}\rightarrow \nu^{\star}\bar{\nu}
\end{equation}
\begin{equation}
e^{-}p\rightarrow \nu^{\star}q(\bar{q})X
\end{equation}
\begin{equation}
pp\rightarrow \nu^{\star}e^+X \qquad\textrm{and\qquad}
pp\rightarrow \nu^{\star}\bar\nu X
\end{equation}

\begin{figure}[h]
\includegraphics{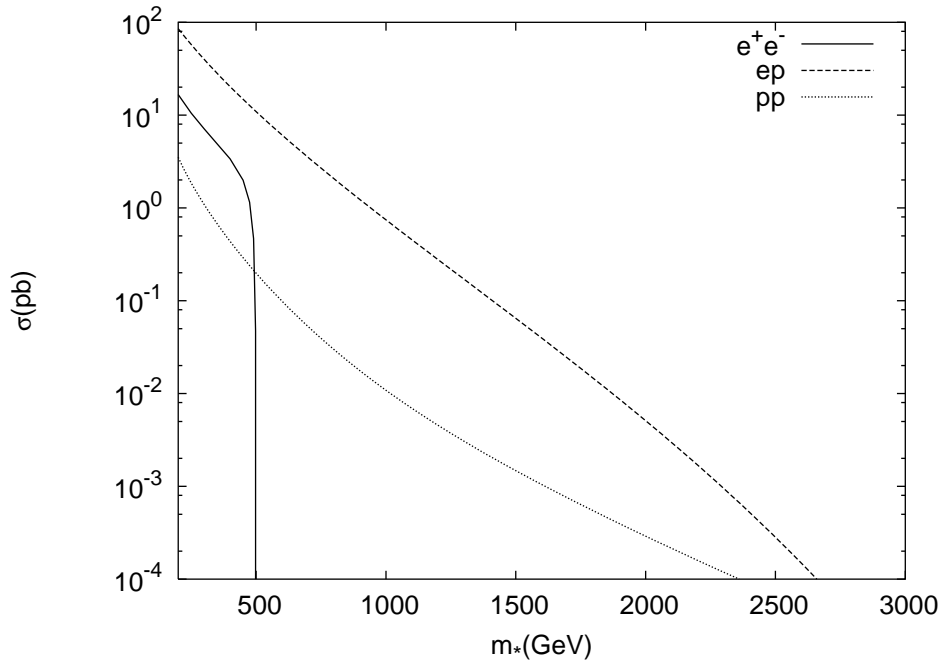}
\caption{The total cross sections for the single production of
excited neutrino with $\Lambda=m_*$ and $f=f^{'}=1$ at
$e^{-}e^{+}$ collider with $\sqrt{s}=0.5$ TeV, $ep$ collider with
$\sqrt{s}=3.74$ TeV and $pp$ collider with $\sqrt{s}=14$ TeV.
\label{fig2}}
\end{figure}

For a comparison of different colliders, the signal cross sections
for the processes given above are presented in Fig. \ref{fig2}
assuming the scale $\Lambda=m_{*}$ and the coupling parameters
$f=f^{'}=1$.

The possible final states for the signal at three different
collisions are given in Table \ref{table1}. As can be seen from
Table \ref{table1} the radiative decays $\nu^{\star}\rightarrow
\nu\gamma$ and neutral currents decays $\nu^{\star}\rightarrow \nu
Z$ will result in larger uncertainty due to the missing transverse
momentum, whereas charged current decays $\nu^{\star}\rightarrow e
W$ leads to a better reconstruction for the signal. In this study,
we consider the subsequent decay of excited neutrino (or
antineutrino) into a W-boson and an electron (or positron).
Therefore, we deal with the process $e^{+}e^{-}\rightarrow
W^+e^{-}\bar\nu$, and subprocesses $e^{-}q(\bar{q'})\rightarrow
W^+e^{-}q(\bar{q})$, $q\bar{q}\rightarrow W^+e^{-}\bar\nu$ and
$q\bar{q'}\rightarrow W^+e^{-}e^+$. The signal and background were
simulated at the parton level by using the program CompHEP 4.2
\cite{Pukhov} (the interference terms between signal and
background processes are taken into account in the program). In
our calculations, we used the parton distribution functions
library CTEQ6L \cite{Lai} with the factorization scale
$Q^{2}=\hat{s}$.
\begin{table}[h]
\caption{Event topologies for the single production of excited
neutrinos with subsequent decay channels at three different type
of collisions.\label{table1}} \footnotesize{
\begin{tabular}[c]{llllllllll}\hline
(sub)process &&& photon channel &&& Z-boson channel &&& W-boson
channel\\
\hline $e^{+}e^{-}\rightarrow\nu^{\star}\overline{\nu}$ &&&
$\gamma\nu\overline{\nu}$ &&&
$Z\nu\overline{\nu}$ &&& $W^{+}e^{-}\overline{\nu}$\\
$q\overline{q}\rightarrow\nu^{\star}\overline{\nu}$ &&&
$\gamma+\not p_{T}$ &&& $2j+\not p_{T}:l^{+}+l^{-}+\not p_{T}:\not
p_{T}$ &&& $\mathbf{2j+e}^{-} +\not p_{T}:l^{+}+e^{-}+\not
p_{T}$\\\hline
$e^{-}q\rightarrow\nu^{\star}q$ &&& $\gamma\nu q$ &&& $Z\nu q$ &&& $W^{+}e^{-}q$\\
&&& $\gamma+j+\not p_{T}$ &&& $3j+\not p_{T}:l^{+}+l^{-}+j+\not
p_{T}:j+\not p_{T}$ &&& $\mathbf{2j+e}^{-}+j:l^{+}+e^{-}+j+\not
p_{T}$\\\hline
$q\overline{q^{^{\prime}}}\rightarrow\nu^{\star}e^{+}$ &&&
$\gamma\nu e^{+}$ &&&
$Z\nu e^{+}$ &&& $W^{+}e^{-}e^{+}$\\
&&& $\gamma+e^{+}+\not p_{T}$ &&& $2j+e^{+}+\not
p_{T}:l^{+}+l^{-}+e^{+} +\not p_{T}:e^{+}+\not p_{T}$ &&&
$\mathbf{2j+e}^{-}+e^{+}:l^{+}+e^{-} +e^{+}+\not p_{T}$\\\hline
\end{tabular} }
\end{table}

\subsection{$e^{+}e^{-}$ Collider}

\begin{figure}
\includegraphics{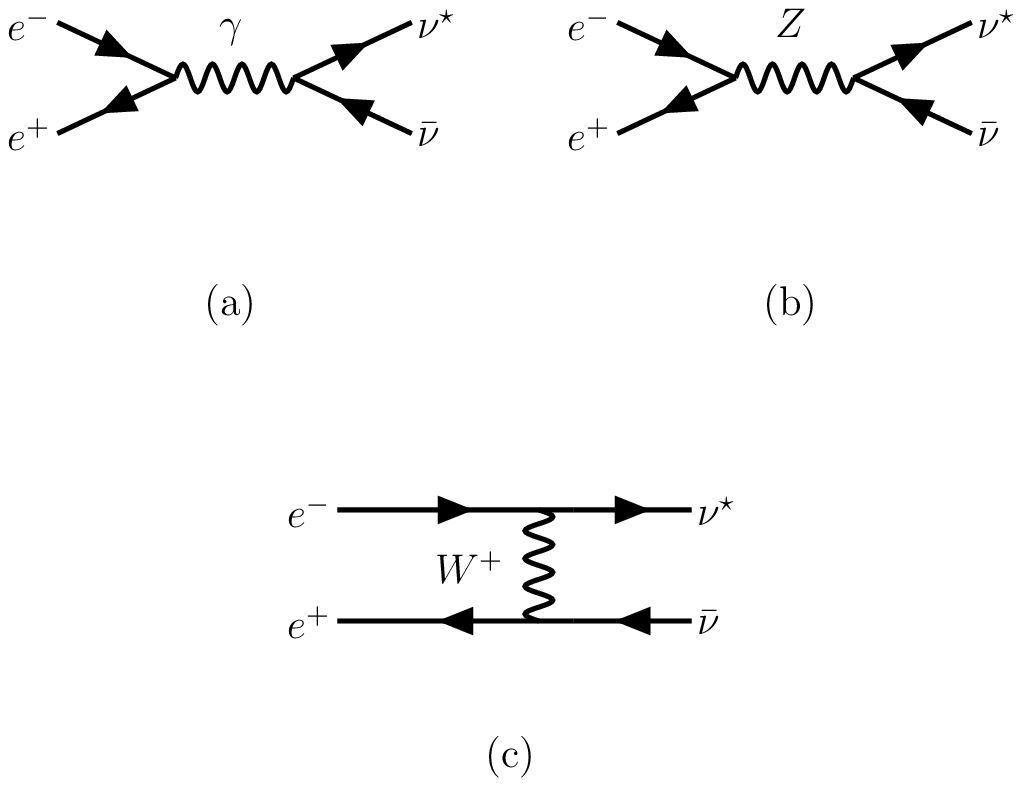}
\caption{Excited neutrino production at $e^{-}e^{+}$ colliders
through the (a,b) $s-$channel and (c) $t-$ channel exchange
diagrams.\label{fig3}}
\end{figure}

High energy electron-positron collisions provide an excellent
environment for the search for excited leptons. We examine the
single production of excited neutrinos ($\nu^{\star}$) at future
$e^{-}e^{+}$ colliders with $\sqrt{s}=500$ GeV, through the
process $e^{-}e^{+}\rightarrow \nu^{\star}\bar\nu\rightarrow
W^+e^{-}\bar\nu$. The Feynman diagram for the process
$e^{-}e^{+}\rightarrow \nu^{\star}\bar\nu$ is shown in Fig.
\ref{fig3}.
\begin{eqnarray}
\frac{d\sigma}{dt} =\frac{g_{e}^{4}}{64\pi s^{2}\Lambda^{2}}
[&-&\frac{2f_{W}^{2}(m_{*}^{2}-s)(m_{*}^{2}-s-t)t}{\sin^{2}\theta_{W}(m_{W}^{2}-t)^{2}}-
\frac{\sqrt{2}(-1+2c_{V})f_{W}f_{Z}st(-m_{*}^{2}+s+t)}
{\cos\theta_{W}(m_{Z}^{2}-s)\sin^{2}\theta_{W}(m_{W}^{2}-t)}\nonumber\\
& -&\frac{4\sqrt{2}f_{\gamma}f_{W}t(-m_{*}^{2}+s+t)}
{\sin\theta_{W}(m_{W}^{2}-t)}
-\frac{4f_{\gamma}^{2}\left[m_{*}^{4}+2t(s+t)-m_{*}^{2}(s+2t)\right]}{s}\nonumber\\
&
+&\frac{f_{Z}^{2}s\left[-(1-2c_{V})^{2}m_{*}^{4}-2(1+4c_{V}^{2})t(s+t)+(1-2c_{V})^{2}m_{*}^{2}(s+2t)\right]}
{4\cos^{2}\theta_{W}(m_{Z}^{2}-s)^{2}\sin^{2}\theta_{W}}\nonumber\\
&-&\frac{2f_{\gamma}f_{Z}\left[(-1+2c_{V})m_{*}^{4}+4c_{V}t(s+t)
-(-1+2c_{V})m_{*}^{2}(s+2t)\right]}
{\cos\theta_{W}(m_{Z}^{2}-s)\sin\theta_{W}} ]
\end{eqnarray}

\begin{figure}
\includegraphics{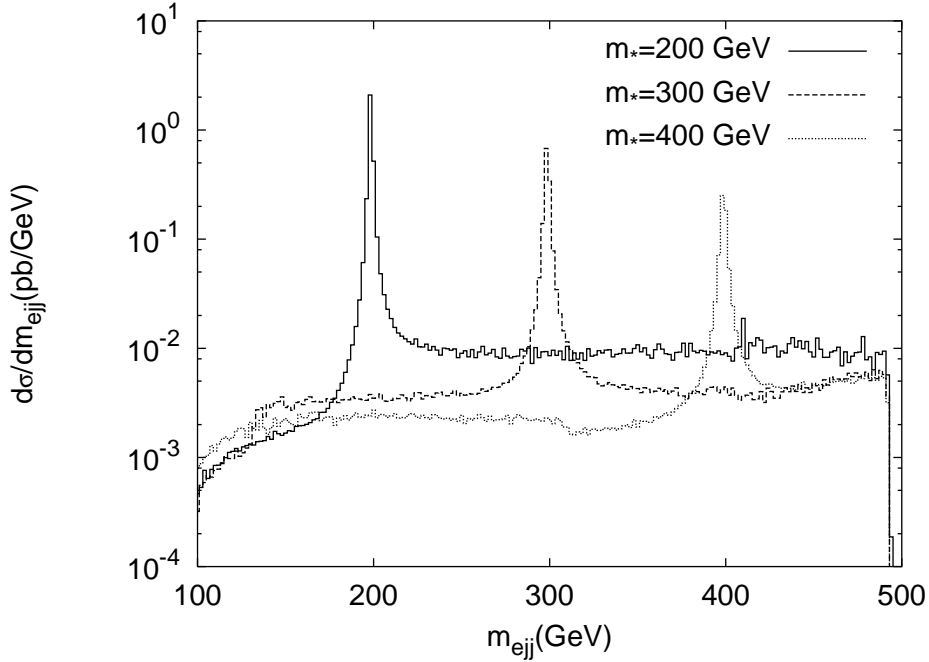}
\caption{Invariant mass $m_{ejj}$ distribution of signal (the
scale $\Lambda=m_{*}$ and the coupling parameters $f=f^{'}=1$) and
the corresponding background at $e^{-}e^{+}$
collider.\label{fig4}}
\end{figure}

We applied the following acceptance cuts to form the signal and
reduce the backgrounds
\begin{equation}
p_{T}^{e^\pm,j}>20\textrm{ GeV}
\end{equation}
\begin{equation}
|\eta_{e^{\pm,j}}|<2.5
\end{equation}
where $p_{T}$ is the transverse momentum of the visible particle
or the missing transverse momentum when a neutrino is produced.
$\eta$ stands for the pseudo-rapidity of the visible particles.
After applying these cuts, the SM background cross section is
found to be $\sigma_B=0.89$ pb. For the W-boson decay we choose
the case $W\to$2 jets. In the case of leptonic decays of W-boson
the final state of the process consists of two neutrinos giving a
large uncertainty in the excited neutrino mass reconstruction. The
$\nu^\star\to W^+e^-$ decay of excited neutrino can be easily
identified since the invariant mass of the $ejj$ system shows a
peak around the excited neutrino mass. Fig. \ref{fig4} shows the
invariant mass $m_{ejj}$ distribution in the reaction
$e^{+}e^{-}\rightarrow W^+e^{-}\bar\nu$ for the SM background and
with the inclusion of an excited neutrino with masses
$m_{\star}=200$ GeV, $m_{\star}=300$ GeV, $m_{\star}=400$ GeV and
parameter $f=f^{'}=1$.

\begin{figure}
\includegraphics{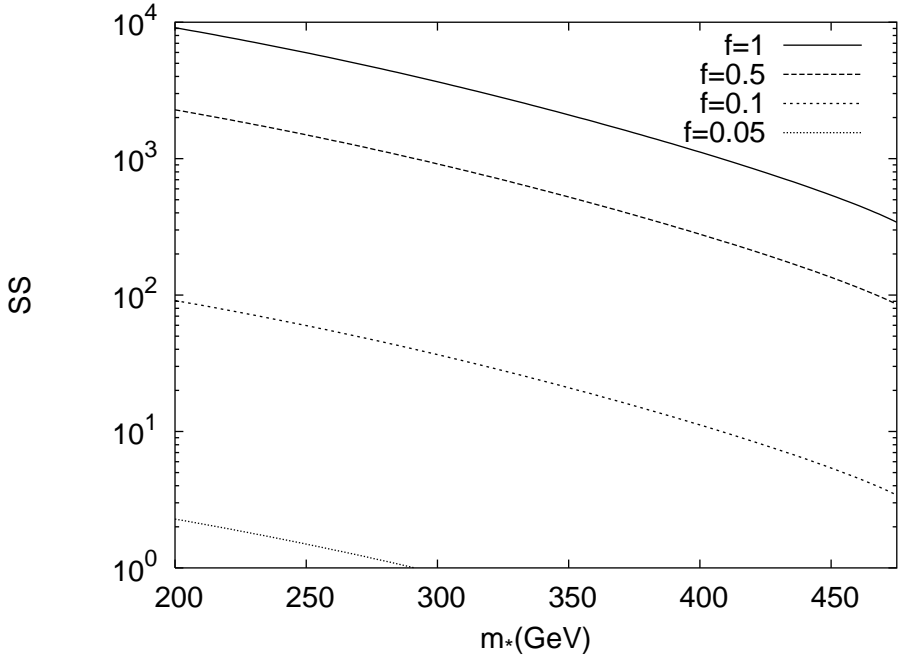}
\caption{Statistical significance depending on the excited
neutrino mass and coupling parameters ($f=f^{'}$) at the LC for
the scale $\Lambda=m_{*}$.\label{fig5}}
\end{figure}

A natural way of extracting the excited neutrino signal, and at
the same time suppressing the SM backgrounds is to impose a cut on
the $ejj$ invariant mass. Therefore, we introduced the cut
\begin{equation}
|m_{ejj}-m_{\star}|<25\textrm{ GeV}
\end{equation}
for considered range of excited neutrino masses. In Table
\ref{table2}, we have presented the signal (for $f=f^{'}=1$) and
background cross sections in $ejj$ invariant mass bins since the
signal is concentrated in a small region proportional to the
invariant mass resolution. In order to examine the potential of
the collider to search for the excited neutrinos, we defined the
statistical significance $SS$ of the signal
\begin{equation}
SS=\frac{\left|\sigma_{S+B}-\sigma_{B}\right|}{\sqrt{\sigma_{B}}}\sqrt{L_{int}}
\end{equation}
where $L_{int}$ is the integrated luminosity of the collider. The
values of $SS$ evaluated at each excited neutrino mass points are
shown in the last column of Table \ref{table2}. As seen from the
Table \ref{table2}, the calculated $SS$ values are higher than 5
up to the center of mass energy of the LC. Single production of
excited neutrinos is feasible up to the center of mass energy of
$e^{+}e^{-}$ collider even with fairly small magnetic transition
couplings to the leptons. For various coupling parameters
$f(=f^{'})$, we give the $SS$ values in Fig. \ref{fig5}.
Concerning the criteria above (SS>5), even for smaller coupling as
$f=f^{\prime}=0.1$ excited neutrinos with masses up to 450 GeV can
be probed at the LC. We could compare the results for the cases
$f=f^{'}=1$ and $f=-f^{'}=1$ to see the effects of photon coupling
$f_\gamma$ in excited neutrino production mechanism. We find
decreases about (15-5)\% in the cross sections for the values of
excited neutrino mass intervals $m_\star=200-400$ GeV. Because of
the high energy of electron/positron beams, initial state
radiation (ISR) can affect the production cross sections. We find
10\% decrease in the signal+background cross section up to the
kinematical energy range of the collision.
\begin{table}
\caption{Statistical significance SS are calculated within
selected mass bin width $\Delta m=50$ GeV for an integral
luminosity of $L_{int}=100$ fb$^{-1}$ at the LC. \label{table2}}
\begin{tabular}{|c|c|c|c|}
\hline
$m_{\star}$(GeV)&
$\sigma_{S+B}$(pb)&
$\sigma_{B}$(pb)&
$SS$\tabularnewline
\hline
200&
$7.51\times10^{0}$&
$6.66\times10^{-2}$&
9120.8\tabularnewline
\hline
250&
$5.04\times10^{0}$&
$6.53\times10^{-2}$&
6156.2\tabularnewline
\hline
300&
$3.34\times10^{0}$&
$7.01\times10^{-2}$&
3905.5\tabularnewline
\hline
350&
$2.15\times10^{0}$&
$8.81\times10^{-2}$&
2196.7\tabularnewline
\hline
400&
$1.45\times10^{0}$&
$1.25\times10^{-1}$&
1185.1\tabularnewline
\hline
450&
$9.53\times10^{-1}$&
$1.75\times10^{-1}$&
588.1\tabularnewline
\hline
475&
$6.11\times10^{-1}$&
$1.68\times10^{-1}$&
341.8\tabularnewline
\hline
\end{tabular}
\end{table}

\subsection{$ep$ Collider}

\begin{figure}[h]
\includegraphics{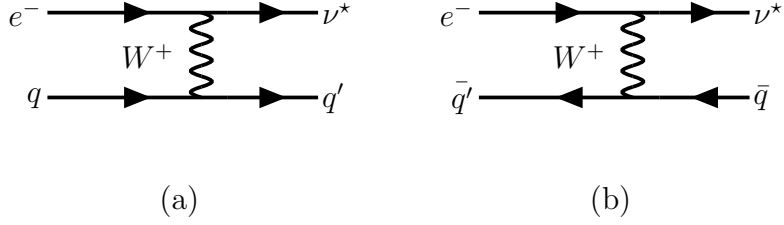}
\caption{Single production of excited neutrinos at $ep$ colliders
through the subprocesses (a) $e^{-}q\rightarrow \nu^{*}q'$ and (b)
$e^{-}\overline{q'}\rightarrow \nu^{*}\overline{q}$.\label{fig6}}
\end{figure}

The magnetic transition couplings of excited neutrino to the
electron allows single production of $\nu^{\star}$ through
$t$-channel $W$ boson exchange. The Feynman diagrams for the
subprocess $e^{-}q\rightarrow \nu^{\star}q'$ and
$e^{-}\overline{q'}\rightarrow \nu^{\star}\overline{q}$ are shown
in Fig. \ref{fig6}.
\begin{eqnarray}
\frac{d\hat{\sigma}}{d\hat{t}}=\frac{f_{W}^{2}\,g_{e}^{4}\,\hat{t}\,
\left(\hat{s}+\hat{t}\right)\,|U_{ij}|^{2}}{32\,\pi\,
\hat{s}{\Lambda}^{2}\,\sin^{2}\theta_{W}\,
{\left(m_{W}^{2}-\hat{t}\right)}^{2}}
\end{eqnarray}

\begin{figure}
\includegraphics{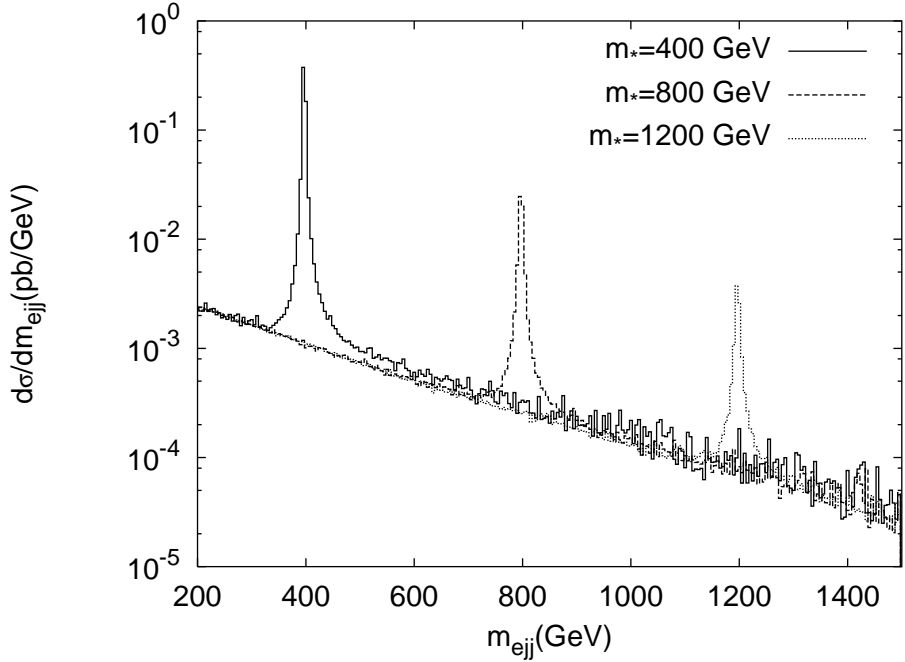}
\caption{Invariant mass $m_{ejj}$ distribution of signal
($\Lambda=m_{*}$ and $f=f^{'}=1$) and background at $e^{-}p$
colliders.\label{fig7}}
\end{figure}

After the acceptance cuts the total SM background cross section is
obtained as $\sigma_{B}=1.11$ pb. Fig. \ref{fig7} shows the
invariant mass $m_{ejj}$ distribution in the reaction
$e^{-}q\rightarrow W^{+}e^-\bar\nu$ for the SM background and the
signal (for $f=f^{'}=1$) with the inclusion of an excited neutrino
with masses $m_{*}=400$ GeV, $m_{*}=800$ GeV and $m_{*}=1200$ GeV.

\begin{figure}[h]
\includegraphics{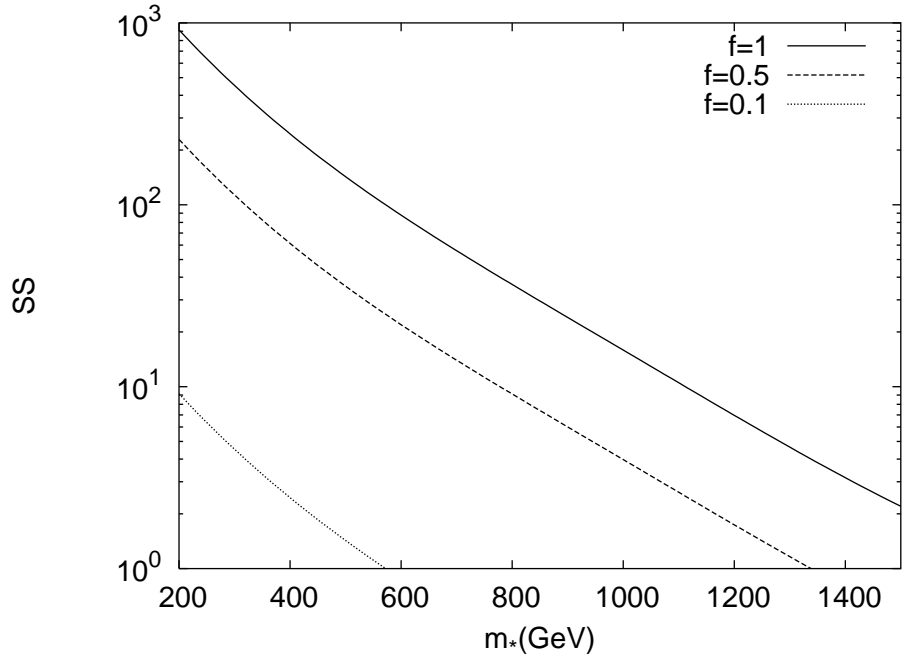}
\caption{Statistical significance depending on the excited
neutrino mass for different coupling parameters ($f=f^{'}$) for
the process $e^{-}p\rightarrow W^+e^{-}q(\bar{q})X$ at the
LC$\otimes$LHC.\label{fig8}}
\end{figure}

In Table \ref{table3}, we present the signal and background cross
sections in $ejj$ invariant mass bins satisfying the condition
$|m_{ejj}-m_{*}|<25\textrm{ GeV}$ for the mass range
$m_{*}=200-1200$ GeV and $|m_{ejj}-m_{*}|<50\textrm{ GeV}$ for
$m_{*}=1200-1500$ GeV. For various coupling parameters
$f(=f^{'})$, we show the mass dependence of the $SS$ in Fig.
\ref{fig8}. As can be seen from Table \ref{table3} LC$\otimes$LHC
can discover excited neutrino in $\nu^{*}\rightarrow W^+e^-$ decay
mode for $f=f^{'}=1$ up to the mass of 1300 GeV. If we take
$f=-f^{'}=1$ and $\lambda=m_\star$, we find almost the same signal
cross sections for higher masses as $m_\star >500$ GeV. However, a
decrease in the cross section about 10\% is obtained at
$m_\star=200$ GeV. The effects from ISR is found to be 5\%
decrease in the signal+background cross section. One can conclude
that the results obtained for the LC$\times$LHC turns out to be at
least an order of magnitude more stringent than the present best
limits coming from the HERA experiments.

\begin{table}
\caption{Statistical significance SS of the excited neutrino
signal with couplings $f=f^{'}=1$ and $\Lambda=m_*$ are calculated
for an integral luminosity of $L_{int}=100$ pb$^{-1}$ at the LC
$\times$ LHC. \label{table3}}
\begin{tabular}{|c|c|c|c|}
\hline $m_{*}$(GeV)& $\sigma_{S+B}$(pb)& $\sigma_{B}$(pb)&
$SS$\tabularnewline \hline 200& $2.96\times10^{1}$&
$1.04\times10^{-1}$& 914.6\tabularnewline \hline 400&
$4.31\times10^{0}$& $4.46\times10^{-2}$& 201.9\tabularnewline
\hline 600& $1.12\times10^{0}$& $2.09\times10^{-2}$&
76.03\tabularnewline \hline 800& $3.62\times10^{-1}$&
$9.71\times10^{-3}$& 35.7\tabularnewline \hline 1000&
$1.32\times10^{-1}$& $5.30\times10^{-3}$& 17.4\tabularnewline
\hline 1200& $5.41\times10^{-2}$& $5.93\times10^{-3}$&
6.25\tabularnewline \hline 1500& $1.35\times10^{-2}$&
$2.50\times10^{-3}$& 2.20\tabularnewline \hline
\end{tabular}
\end{table}

\subsection{$pp$ Collider}

\begin{figure}[h]
\includegraphics{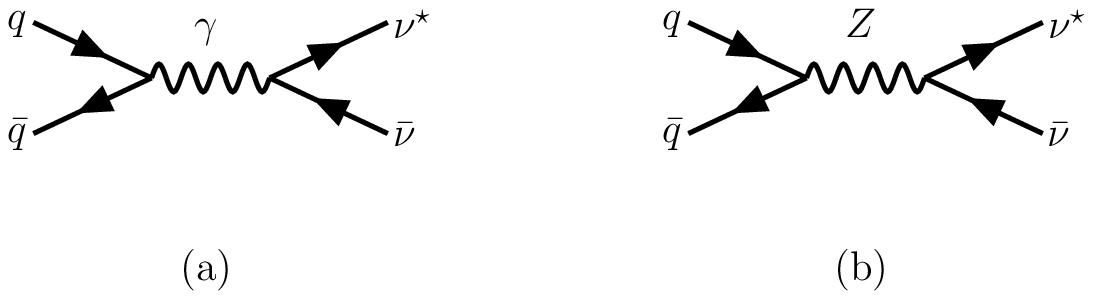}\\
\vspace{1cm}
\includegraphics{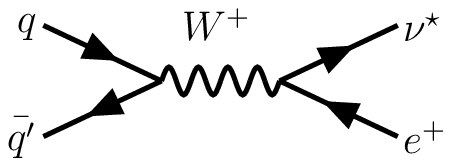}\\
\centerline{(c)} \caption{Excited neutrino production via the
(a,b) photon and $Z-$boson exchange, and (c) $W^{+}$ boson
exchange diagrams at hadron colliders.\label{fig9}}
\end{figure}

At the LHC, the single production of excited neutrinos takes place
through the subprocesses $q\bar{q}\rightarrow
Z^*/\gamma^*\rightarrow \nu^{*}\bar\nu\rightarrow W^+e^{-}\bar\nu$
and $q\bar{q}^{'}\rightarrow W^{+}\rightarrow \nu^{*}e\rightarrow
W^+e^-e^+$ via the Drell-Yan mechanism. The diagrams related to
these subprocesses are shown in Fig. \ref{fig9}. After the
acceptance cuts the total SM background cross sections are
obtained as $\sigma_{B}=2.73$ pb for the process $pp\rightarrow
W^{+}e^{-}\bar\nu X$ and $\sigma_{B}=0.19$ pb for $pp\rightarrow
W^+e^-e^{+}X$.
\begin{eqnarray}
\frac{d\widehat{\sigma}}{d\widehat{t}}(q\overline{q}
\rightarrow\nu^{*}\overline{\nu})&=&\frac{g_{e}^{4}}{1024\pi\widehat{s}^{2}\Lambda^{2}}[
-\frac{64f_{\gamma}^{2}Q_{q}^{2}[m_{*}^{2}
+2\widehat{t}(\widehat{s}+\widehat{t})-m_{*}^{2}(\widehat{s}+2\widehat{t})]}{\widehat{s}} \\
&+&\frac{64f_{\gamma}f_{Z}Q_{q}[c_{A}m_{*}^{2}(m_{*}^{2}-\widehat{s}-2\widehat{t})+c_{V}[m_{*}^{4}
+2\widehat{t}(\widehat{s}+\widehat{t})-m_{*}^{2}(\widehat{s}+2\widehat{t})]]}{\cos\theta_{W}\sin\theta_{W}
(m_{Z}^{2}-\widehat{s})} \nonumber \\
&+&\frac{f_{Z}^{2}s[2c_{A}c_{V}m_{*}^{2}(-m_{*}^{2}+\widehat{s}+2\widehat{t})-(c_{A}^{2}+c_{V}^{2})
(m_{*}^{4}+2\widehat{t}(\widehat{s}+\widehat{t})-m_{*}^{2}(\widehat{s}+2\widehat{t}))]}{\cos^{2}\theta_{W}
\sin^{2}\theta_{W}(m_{Z}^{2}-\widehat{s})^{2}}]\nonumber
\end{eqnarray}
\begin{eqnarray}
\frac{d\widehat{\sigma}}{d\widehat{t}}(q\overline{q^{\prime}}\rightarrow\nu^{*}l^{+})=\frac{f_{W}^{2}g_{e}^{4}(m_{*}^{2}-\widehat{s}-\widehat{t})(-m_{*}^{2}+\widehat{t})|U_{ij}|^{2}}{64\pi\widehat{s}\Lambda^{2}\sin^{2}\theta_{W}(m_{W}^{2}-\widehat{s})^{2}}
\end{eqnarray}

\begin{figure}[h]
\includegraphics{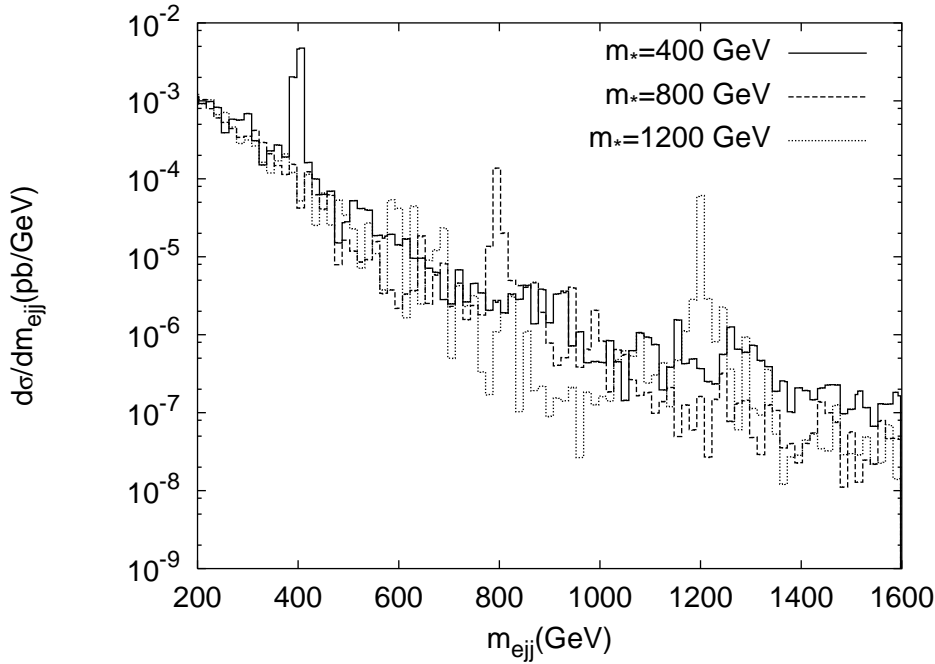}
\caption{Invariant mass $m_{ejj}$ distribution of signal and
background for the processes  $pp\rightarrow W^+e^{-}e^{+} X$ at
$pp$ collider (LHC).\label{fig10}}
\end{figure}

\begin{figure}[h]
\includegraphics{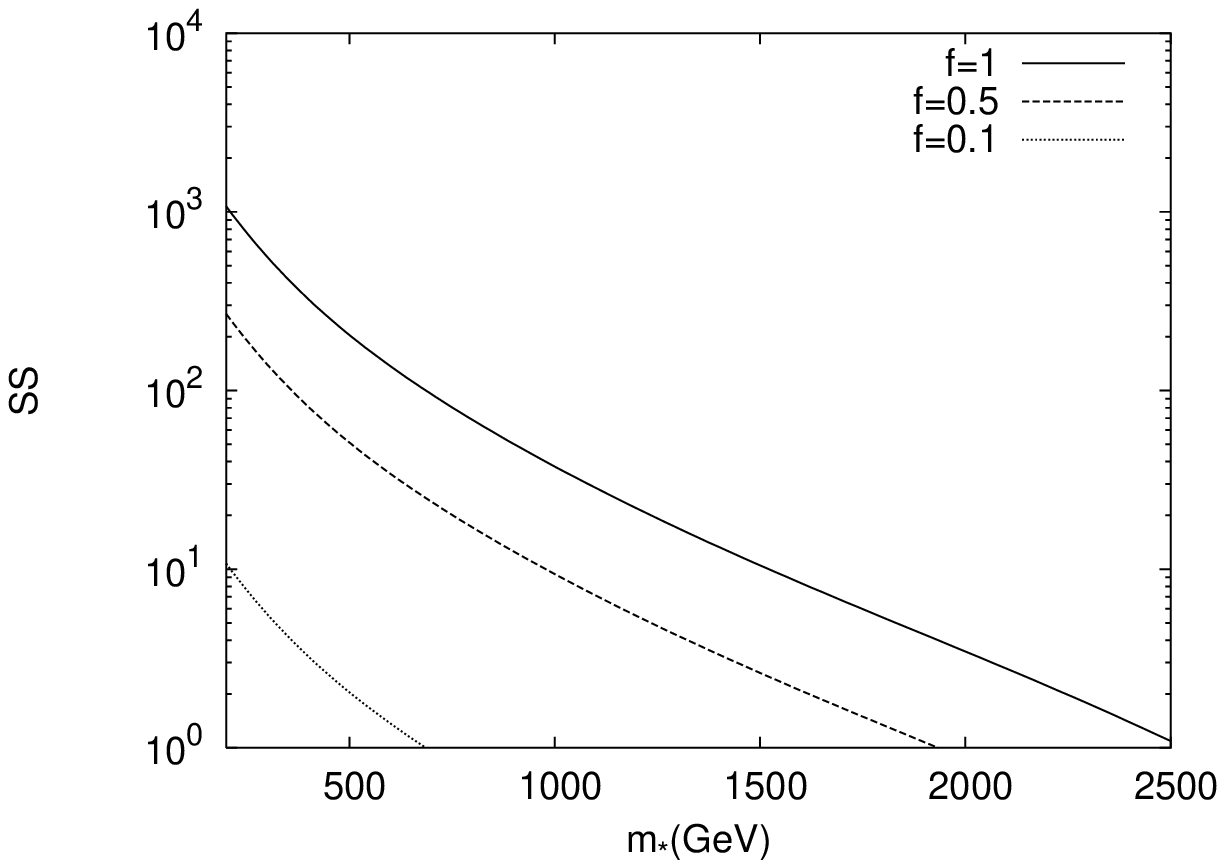}
\caption{Statistical significance depending on the excited
neutrino mass for the process $pp\rightarrow W^+e^{-}e^{+} X$ with
different couplings $f(=f')$ at the LHC.\label{fig11}}
\end{figure}

Fig. \ref{fig10} shows the invariant mass $m_{ejj}$ distributions
in the reactions $pp\rightarrow W^+e^{-}e^+ X$ for the SM
background and the signal (for $f=f^{'}=1$) with masses of
$m_{*}=400$ GeV, $m_{*}=800$ GeV and $m_{*}=1200$ GeV. In Table
\ref{table4}, we present the signal and background cross sections
in $ejj$ invariant mass bins satisfying the condition
$|m_{ejj}-m_{*}|<25\textrm{ GeV}$ for excited neutrino mass
$m_{*}=200-1200$ GeV and $|m_{ejj}-m_{*}|<50\textrm{ GeV}$ for
$m_{*}=1200-2000$ GeV. Statistical significance SS are shown in
Fig. \ref{fig11} for $pp\rightarrow W^+e^{-}e^{+} X$ process with
different couplings $f(=f')$. For the parameters $f=-f^{'}=1$, we
find a maximum decrease about 5\% in the signal cross sections for
the considered range of excited neutrino masses. Relatively weak
limits on the masses and couplings are obtained for the process
$pp\rightarrow W^+e^{-}\bar\nu X$. One can conclude that the LHC
will be able to extend considerably the range of excited neutrino
masses up to about 1.85 TeV.
\begin{table}
\caption{The cross sections of signal and background, and the
statistical significance SS of the excited neutrino signal with
couplings $f=f^{'}=1$ and $\Lambda=m_*$ are calculated for an
integral luminosity of $L=100$ fb$^{-1}$ at the LHC.
\label{table4}}
\begin{tabular}{|c|c|c|c|c|c|c|}
\hline process$\rightarrow$& \multicolumn{3}{c|}{$pp\rightarrow
e^{-}W^{+}e^{+}X$}& \multicolumn{3}{c|}{$pp\rightarrow
e^{-}W^{+}\nu X$}\tabularnewline \cline{1-1} \cline{5-7} \hline
$m_{*}$(GeV)& $\sigma_{S+B}$(pb)& $\sigma_{B}$(pb)& $SS$&
$\sigma_{S+B}$(pb)& $\sigma_{B}$(pb)& $SS$\tabularnewline \hline
200& $7.70\times10^{-1}$& $4.98\times10^{-2}$& 1074.0&
$7.34\times10^{-1}$& $4.65\times10^{-1}$& 124.7\tabularnewline
\hline 400& $7.10\times10^{-2}$& $7.16\times10^{-3}$& 256.9&
$5.90\times10^{-2}$& $4.67\times10^{-2}$& 17.9\tabularnewline
\hline 600& $1.53\times10^{-2}$& $1.61\times10^{-3}$& 116.3&
$1.20\times10^{-2}$& $8.33\times10^{-3}$& 12.7\tabularnewline
\hline 800& $4.57\times10^{-3}$& $5.02\times10^{-4}$& 62.3&
$3.03\times10^{-3}$& $2.45\times10^{-3}$& 3.7\tabularnewline
\hline 1200& $7.50\times10^{-4}$& $1.64\times10^{-4}$& 17.1&
$5.49\times10^{-4}$& $7.13\times10^{-4}$& 1.9\tabularnewline
\hline 1600& $1.65\times10^{-4}$& $3.66\times10^{-5}$& 7.9& $-$&
$-$& $-$\tabularnewline \hline 2000& $4.15\times10^{-5}$&
$9.96\times10^{-6}$& 3.8& $-$& $-$& $-$\tabularnewline \hline
\end{tabular}
\end{table}

\section{Conclusion}

In our analysis, we assumed that the excited neutrino interact
with the SM particles via the effective Lagrangian (1). We
restrict ourselves to the gauge interactions since the aim of this
paper is to compare the potential of three types of colliders
within the similar sets of cuts. Our results coincide with
\cite{Boudjema} where similar analysis was performed for the LC,
and essentially coincide with \cite{Belyaev} where the mass range
accessible with LHC was obtained. Finally, our analysis show that
for relatively small masses ($m_*<450$ GeV) LC is more promising
than LC$\otimes$LHC and the LHC for the processes considered. The
LHC can probe higher masses of excited neutrinos and smaller
couplings than the LC and LC$\times$LHC can do.

We may conclude from Table \ref{table4} that the $W^+e^-\nu$ final
state in $e^+e^{-}$ and $pp$ collisions has both excited neutrino
and excited electron contribution. However, $W^+e^-q$ and
$W^+e^-e^+$ final state in $e^{-}p$ and $pp$ collisions isolate
the excited neutrino contribution. Focusing on the $2j$+opposite
sign leptons signal in $pp$ collision and $3j$+a lepton signal (by
requiring a peak in the $ejj$ invariant mass distribution and
reconstructing $W$-mass from two jets) excited neutrino can be
probed at the detectors built on the future colliders.

We give the realistic estimates for excited neutrino signal and
the corresponding background at three-type of colliders with the
availability of higher center of mass energies and higher
luminosities. Since the cross section for the signal is
proportional to $1/\Lambda^{2}$, various choices of the $\Lambda$,
i.e., in this study we have choosen $\Lambda=m_{*}$, will lead to
the changes in the cross sections as $(m_{*}/\Lambda)^{2}.$ For
$\Lambda=1$ TeV, we need to multiply the signal cross sections by
a factor {[}$m_{*}$(TeV){]}$^{2}$ at every mass values of excited
neutrinos. Our analysis show that for $f=f^{'}=1$ the LC can
discover excited neutrino in $\nu^{*}\rightarrow W^+e^-$ decay
mode up to the kinematical limit, while LC$\otimes$LHC and LHC can
reach much higher mass values, namely 1300 GeV and 1850 GeV,
respectively. For $f=f^{'}=0.1$ discover limits are: 450 GeV at
LC, 275 GeV at LC$\otimes$LHC and 350 GeV at LHC.

\begin{acknowledgments}
This work is partially supported by Turkish State Planning
Committee under the Grants No 2003K120190.
\end{acknowledgments}

\end{document}